# ANDROID BASED SECURITY AND HOME AUTOMATION SYSTEM

Sadeque Reza Khan[1] and Farzana Sultana Dristy[2]

[1]Department of Information and Communication Engineering, Chosun University, Korea

[2]Department of Computer Science and Engineering, Varendra Universty, Bangladesh

## ABSTRACT

*The smart mobile terminal operator platform Android is getting popular all over the world with its wide variety of applications and enormous use in numerous spheres of our daily life. Considering the fact of increasing demand of home security and automation, an Android based control system is presented in this paper where the proposed system can maintain the security of home main entrance and also the car door lock. Another important feature of the designed system is that it can control the overall appliances in a room. The mobile to security system or home automation system interface is established through Bluetooth. The hardware part is designed with the PIC microcontroller.*

## 1. INTRODUCTION

Developed in Linux kernel, Android platform is composed of operating system, user interface and application components which allow developer freedom access and modify the source code [1]. So Android is providing a free platform to the developers with numerous facilities to generate new applications in a rapid rate.

Wireless technologies are becoming more popular around the world and for a short distance communication, embedded Bluetooth technology can form a network of digital devices, in which the appliances and devices can communicate with each other. Bluetooth technology is the gift for the modern home automation. Operated over 2.4 GHz frequency, Bluetooth technology can link digital devices within a range of 10m to 100m at the speed of up to 3 Mbps depending on the Bluetooth device class [2], [3].

This paper presents an Android application which can be interfaced with three different systems, home security system [4], [5], home automation system and car lock system, respectively, using Bluetooth communication protocol. This arrangement facilitates a user with multiple password based security which is increasing the daily life safety. This architecture also makes ease of controlling home appliances through a simple user interface. The Android SDK tool is used for application development and the controller part is designed with PIC microcontroller and Flowcode ver.5 compiler.

## 2. BLUETOOTH COMMUNICATION WITH ANDROID

The Android-Bluetooth communication process flow [6], [7] is shown in figure 1.

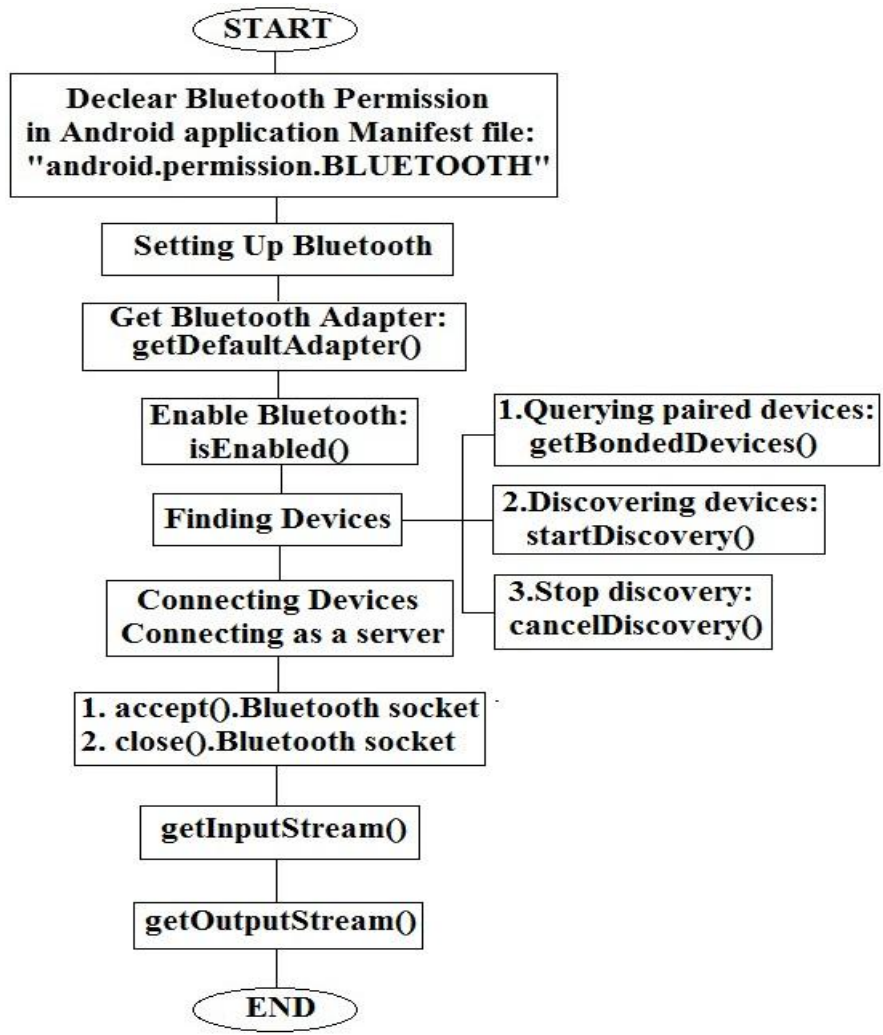

Fig. 1. Android-Bluetooth Communication Flow

To start communication with Bluetooth, Android first seeks permission which is required for requesting a connection, accepting a connection, and transferring data. To discover local Bluetooth devices this permission is required. The **Bluetooth Adapter** is required for all Bluetooth activity and it is called by the static **getDefaultAdapter()** method. This returns a **Bluetooth Adapter** that represents the device's own Bluetooth adapter (the Bluetooth radio). If **getDefaultAdapter()** results null, then the Bluetooth is not supported by the device. Now **isEnabled** () is called to check whether Bluetooth is currently enabled. The false output coming from this method makes the Bluetooth disable and to enable the Bluetooth again, **startActivityForResult**() is called with the **ACTION_REQUEST_ENABLE** action intent which will issue a request to enable Bluetooth through the system settings. Although the next operation is to perform device discovery, before that it is better to query the set of paired

16

devices to see if the desired device is already known by calling **getBondedDevices()**. **startDiscovery()** method is used to start discovering devices in the surrounding. This asynchronous method and can immediately return with a Boolean, indicating whether discovery has successfully started. Performing device discovery is a power consuming procedure. With **cancelDiscovery()** method the search for another device can be stopped immediately, once a device is found to connect. For connection establishment between two devices, one must act as a server by holding an open **BluetoothServerSocket**. The purpose of the server socket is to listen for incoming connection requests and when one is accepted, provide a connected **BluetoothSocket**. Now connection requests can be listened by calling **accept()** method which is actually a blocking call that will return when either a connection has been accepted or an exception has occurred. The acceptance of a connection request is depended on a valid UUID (Universally Unique Identifier) matching, the one registered with this listening server socket and for a successful matching, **accept()** will return a connected **BluetoothSocket**. Now for avoiding additional connections, call **close()** method which releases the server socket and all its resources, but does not close the connected **BluetoothSocket** that's been returned by **accept()**. As the **accept()** method is a blocking, it should not be executed in the main activity UI thread because it will prevent any other interaction with the application. Using the **BluetoothSocket**, transferring arbitrary data can be established by calling the **InputStream** and **OutputStream** which handle transmissions through the socket, via **getInputStream()** and **getOutputStream()**, respectively.

## 3. PRIMARY SECURITY STAGE

### 3.1. Software Section

The designed android application is protected with the particular user name and password in the initial level which is shown in figure 2. For wrong user name or password the system generates a toaster "Invalid User Name or Password". For a correct name and password the system provides access to the main software page. Before entering to the main thread the Android asks permission for turning on the Bluetooth socket in the mobile phone.

To connect with the external Bluetooth module user has to press connect which will appear from the menu tab of the Android mobile phone. If the connection is established properly a confirmation toaster will appear on the screen. Proper UUID and MAC (Media Access Control) address must be provided to establish the connection between the Bluetooth module of the user mobile phone and the Bluetooth module of the particular device.

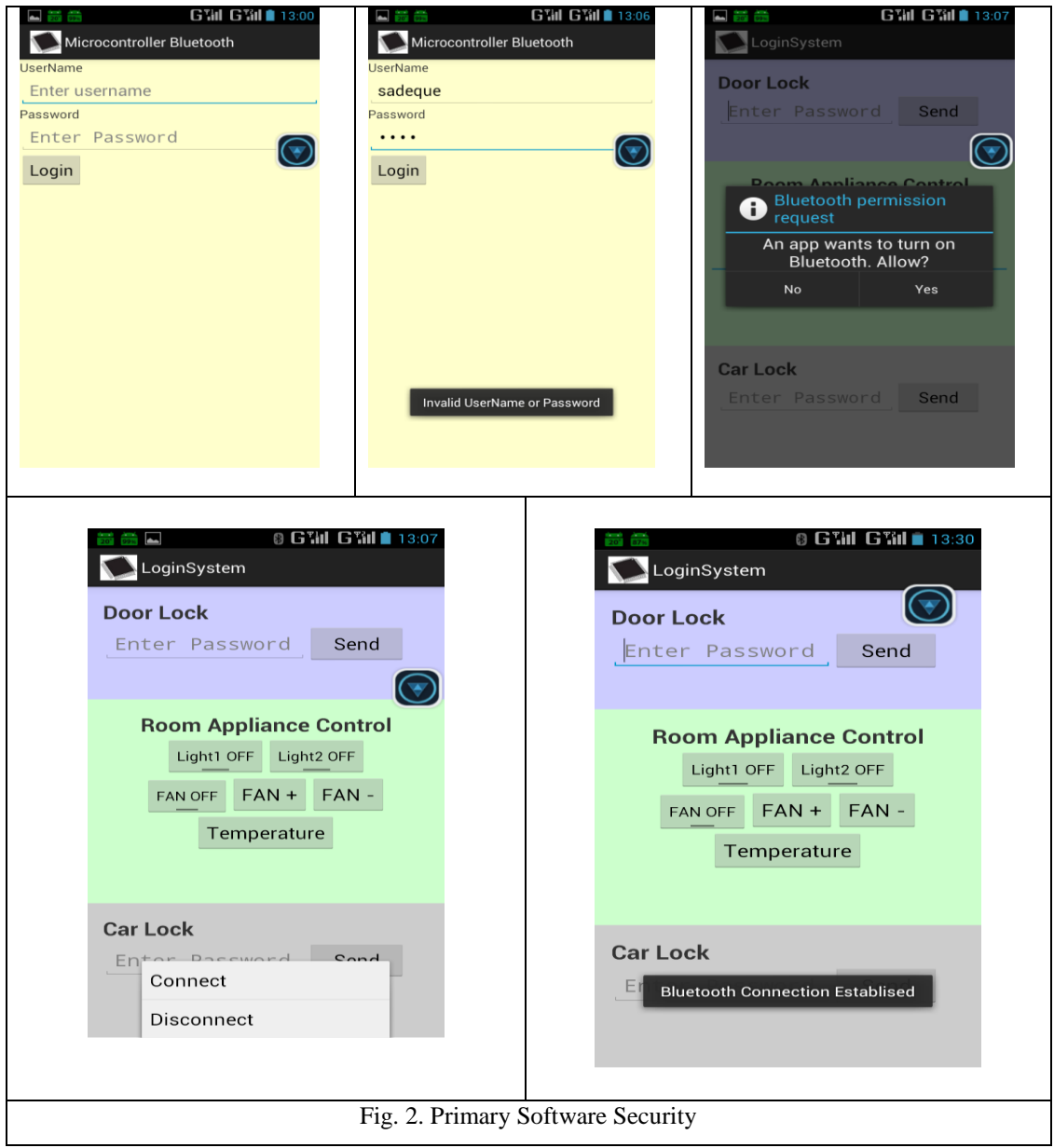

Fig. 2. Primary Software Security

## 3.2. Hardware Section

The hardware section of the primary security is shown in figure 3. To access the main hardware the user has to provide a password which will open the Bluetooth module that is connected in the hardware section. As a central controller PIC microcontroller 16F877A is used. HC-06 Bluetooth module is used here for the purpose of communication with the android system. This module is a serial device and is communicated with central controller by using RS-232 protocol. A 4X3 keypad and a LCD are used for providing password and for checking the system status respectively. If wrong password is pressed three consecutive times, the system sends an SMS (Short Message Service) to the house owner and another one to the nearby police station and the whole system also get collapsed automatically which is not possible to repair without an authorized person. A secuity alarm also gets activated. This features the security of the user's residence and it is implemented with a GSM (Global System for Mobile) Modem. The LCD output is provided in figure 4.

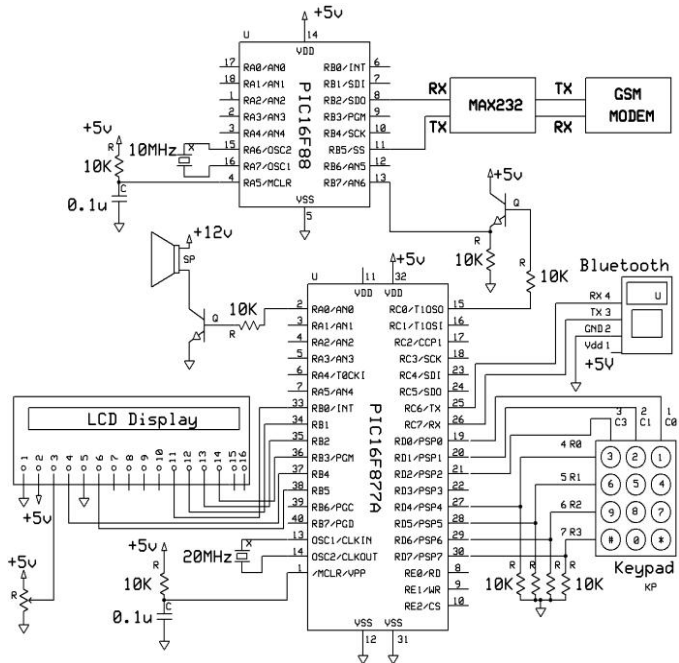
Fig. 3. Hardware Section for Primary Security

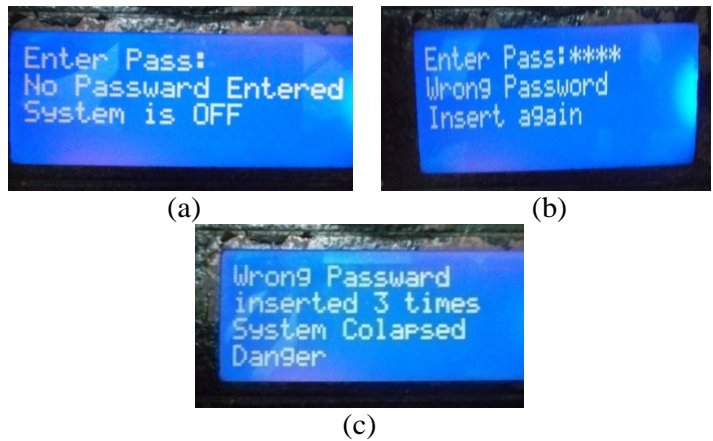
Fig. 4. LCD Output of the Security System

## 4. DOOR LOCK SYSTEM

Keypad is used to provide proper password to the door lock system to turn on the Bluetooth module that is connected to the door and it is a significant feature of the primary security stage. For multiple wrong entries the system will be collapsed automatically and it will send SMS to two particular phone numbers (house owner and police station) that is shown in figure 6. The system will also enable alarm at the same time. Figure 5 shows the door lock system.

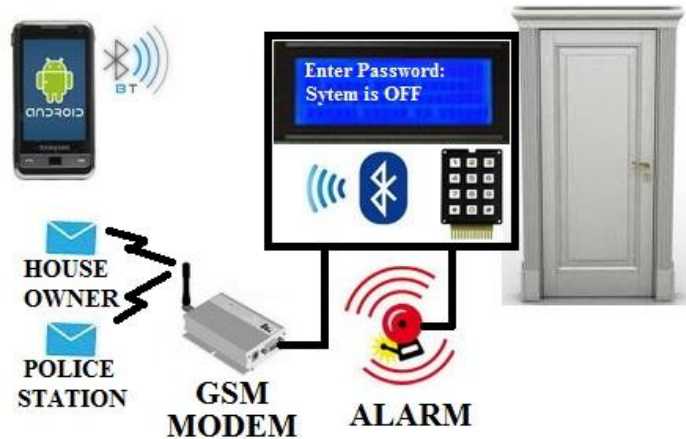

Fig. 5. Door Lock System

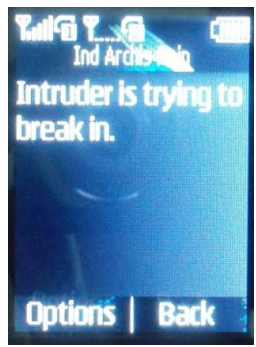

Fig. 6. Received SMS

Figure 7 shows the door lock section of the software. After typing the preset password while the user will press the send button the **outStream.write(Buffer)** method will send the password through Bluetooth socket. At the moment of providing the password, it should be stored in a predefined buffer. Once the password is sent a confirmation twister will appear on the screen.

## 5. HOME AUTOMATION

Figure 8 shows the home appliance control system using the Android and the Bluetooth module. The designed system can control two lights and a fan in a particular room. Figure 9 shows the schematic of the home automation, where PIC16F876A is central controller. A zero crossing detector is connected in RB0 (Interrupt) pin to identify the zero transition of the phase and control the fan speed accordingly through RB5 pin. RB3 and RB4 pins are used to control two lights in a room. There is also a temperature sensor (DS18B20) connected to RA0 pin to transmit room temperature through the Bluetooth module to Android device.

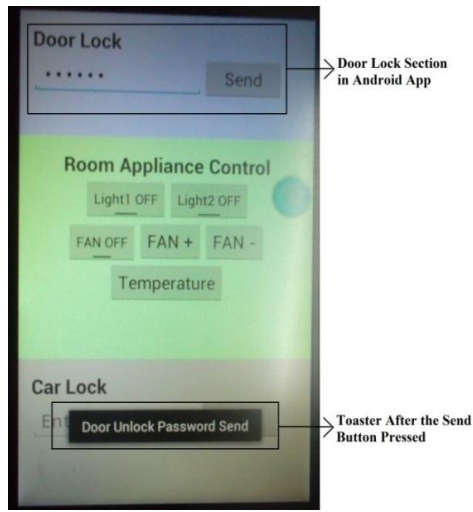

Fig. 7. Android App Door Lock Section

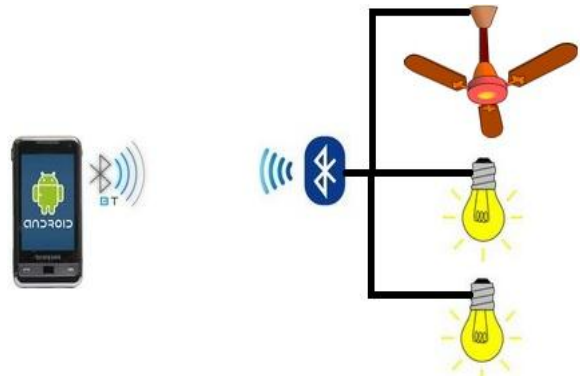

Fig. 8. Home Appliance Control

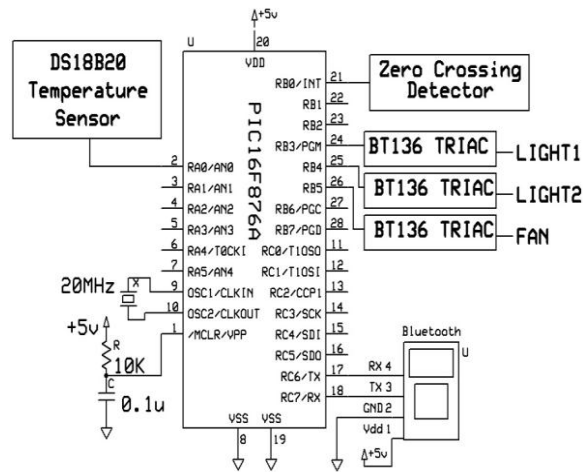

Fig. 9. Schematic for Home Appliance System

Figure 10-13 are showing home automation part of the designed Android application. Figure 10 is showing that, both the light buttons are pressed and it is represented by the title in the buttons as "Light1 On" and "Light2 On" messages. In figure 11 the fan button is pressed and the title is showing "FAN On". Also it is showing a toaster of "Speed Increasing" as "FAN+" button is

21

pressed. Again, figure 12 shows the toaster "Speed Decreasing" as "FAN-" button is pressed. Figure 13 shows the room temperature that is received from the mounted hardware in the room. By pressing the "Temperature" button user can check the room temperature any moment of time. Receiving data through the Bluetooth socket is a bit lengthy process than data sending. Data is received using **btSocket.getInputStream()** method and initially put in to a string variable and further processing is done by Android string manipulation method. Figure 14 is showing the LCD output of the home automation system.

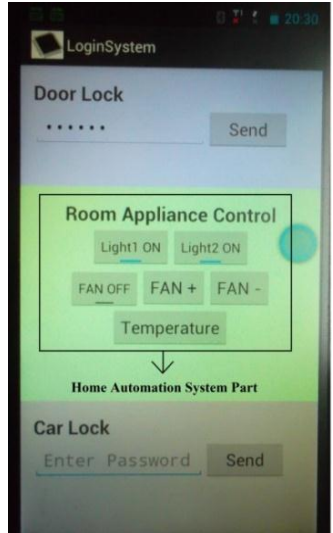
Fig. 10. Home Automation Part in the Android App

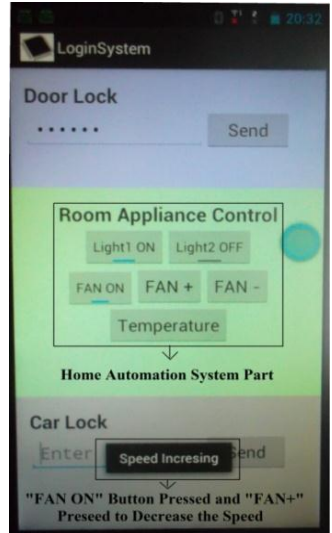
Fig. 11. FAN Speed Increasing

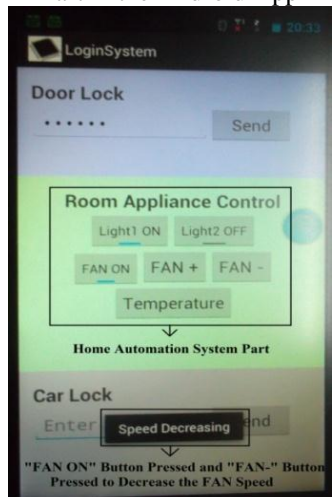
Fig. 12. FAN Speed decreasing

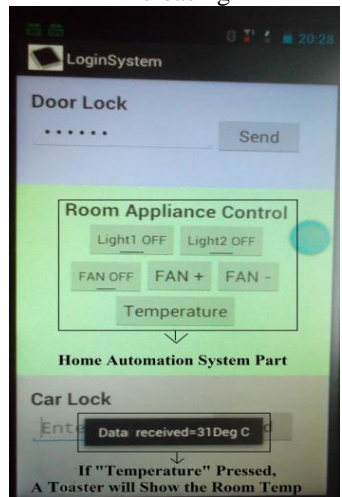
Fig. 13. Room Temperature Monitoring Using the Application

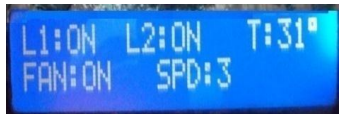
Fig. 14. LCD Output

## 6. CAR LOCK

Figure 15 shows the designed car door lock system and Figure 16 is the schematic of the proposed system. The car door locks are activated by DC solenoids. Depending on the direction of current flow the armature moves either "forward" or "reverse" direction. Normally the relays connect the passenger's door lock actuators to ground on both the sides. When the Android application sends a password through its Bluetooth socket to the Bluetooth module on the car, the PIC16F876A MCU activates Relay 1 (RL1), which puts +12V at actuator connection, causing it to lock. For another password from the Android Bluetooth, the MCU activates Relay 2 (RL2), which puts +12V at the actuator, causing it to unlock. Figure 17 shows the Android application part of the car door lock system where a twister is appeared for sending a door unlock password.

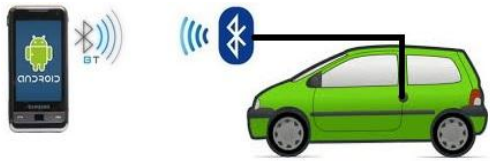

Fig. 15. Car Lock System

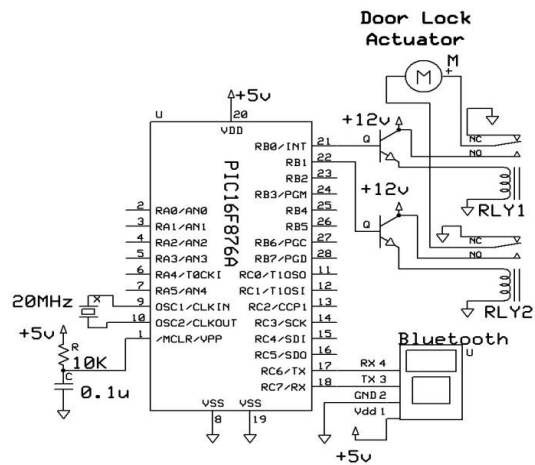

Fig. 16. Schematic of Car Door Lock System

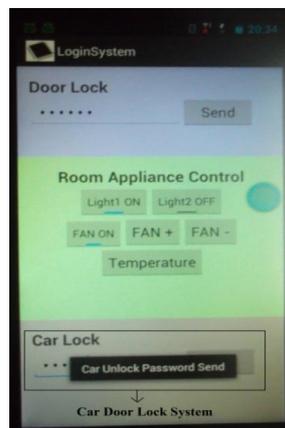

Fig. 17. Android application of Car Door Lock System

23

## 7. CONCLUSIONS

In this paper a security interface and home automation system is presented using an Android mobile device. It is a short range system that is simple to use and easy to interface. Multiple layered passwords are implemented to make the whole system versatile and trustworthy for the users. Home automation system part is also very flexible and user friendly. In future this proposed system can also be expanded to multiple doors and windows and more home appliances can be integrated with the system.

## Authors

**Sadeque Reza Khan**

Sadeque Reza Khan received B.Sc. degree in Electronics and Telecommunication Engineering from University of Liberal Arts Bangladesh and He complered his M.Tech in VLSI Design from National Institute of Technology Kernataka (NITK), India. Currently he is doing Ph.D at SoC Design laboratory, Chosun University, Korea. His research interest includes VLSI, Microelectronics, Control System Designing and Embedded System Designing

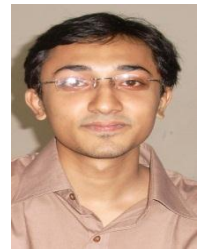

**Farzana Sultana Dristy**

Farzana Sultana Dristy is currently continuing her B.Sc. degree in Computer Science and Engineering at Varendra University, Bangladesh. She is involved in the research fields of Embedded System Designing and Robotics.

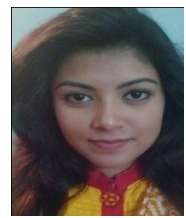